# A Flexible Mixed Integer Programming framework for Nurse Scheduling

Murphy Choy
Michelle Cheong

## Abstract

In this paper, a nurse-scheduling model is developed using mixed integer programming model. It is deployed to a general care ward to replace and automate the current manual approach for scheduling. The developed model differs from other similar studies in that it optimizes both hospital's requirement as well as nurse preferences by allowing flexibility in the transfer of nurses from different duties. The model also incorporated additional policies which are part of the hospital's requirement but not part of the legislations. Hospital's key primary mission is to ensure continuous ward care service with appropriate number of nursing staffs and the right mix of nursing skills. The planning and scheduling is done to avoid additional non essential cost for hospital. Nurses' preferences are taken into considerations such as the number of night shift and consecutive rest days. We will also reformulate problems from another paper which considers the penalty objective using the model but without the flexible components. The models are built using AIMMS which solves the problem in very short amount of time.

## Introduction

Nurse scheduling is a multi-objective problem. Systematic approach for nurse allocation is needed to ensure continuous and adequate level of patient care services while maintaining the legislative requirements as well as internal policies. This problem becomes complex when addition factors such as patient admission, nurse qualifications or license to practice, type of disease as well as unforeseen accidents. The personal needs of the nurses such as vacation or work shift preferences add a new dimension to the scheduling problem. The need to balance all the various dimension of the problem makes nurse scheduling a particularly daunting manual task.

The schedule should specify the day-to-day shift assignments for every nurse in a specified time horizon which satisfy the requirements. It should attempt to be fair in terms of the distribution of shifts and satisfy the needs of the nurses. There are various grades of nurses ranging from registered nurse to junior nurse. Some the nurses might be trained to manage certain medical conditions or skilled in certain area such as intensive care. Due to the varied trainings and specializations, certain type of nurses has to be staffed for wards requiring those skills. These varied conditions cause manual nurse scheduling to consume a significant amount of time. Even when the schedule has been planned manually, it does not necessarily guarantee the fairness of distribution of work such as the number of night shifts or weekend shifts. While the nurses might have indicated their preferences, the planner might not have taken all these into consideration resulting in poorly designed schedules which has to be modified by the nurses swapping duties or working under undesired conditions. Occasionally, the plans did not attempt to efficiently utilize the manpower properly.

The flexible mixed integer programming model attempts to develop a simple approach to the nurse scheduling problem by modifying the nursing scheduling problem into a linear form which can be solved by a variety of solvers in Excel. The model also attempts to enforce fairness and incorporate the nurses' preferences to maximize the perceived equality and morale.

There are several sections to the paper. In the next section, we will review the existing literature. The mixed integer model will be covered in section 3 where nursing policies, legislation, human resource policies and nursing preferences will be described. The various policies and constraints will then be constructed as mathematical forms which can then be used in the formulation. In Section 4, we will discuss the results, conclusion and future directions in research.

**Literature review**

Nursing scheduling has been widely studied since 1960s. Prior to the development of mathematical programming, most nursing scheduling approaches were based on cyclical modeling. Cyclic models consist of regular patterns which can be rotated across multiple time periods. The pattern will only repeat after one cycle. These types of models are highly repetitive and regular. Even though such models are considered to be fair in terms of distribution of work, the modeling process ignores the preferences of the nurses. Howell's approach (1966) provides the first cyclical scheduling approach which takes into considerations the behavior and preferences of the individual nurse. Subsequently, nurse scheduling began to adopt heuristic models which are able to consider all the requirements at the planning stage (Maier-Rothe and Wolfe, 1973; Isken and Hancock, 1991). This enables the models to attempt satisfying all the requirements. The development of mathematical programming also gave rise to various approaches to solving the nurse scheduling problem especially the non-cyclical problem (Harmeier, 1991; Ozkarahan, 1989; Ozkarahan, 1991; Tobon, 1984; Warner, 1976b).

Literature involving exact algorithms usually consider simplified scenario such as cyclic models which are not realistic while heuristic approaches can result in serious solution feasibility issues. Such issues can lead to labor grievance. In certain situations, the solution derived heuristically may be suboptimal. Most heuristic solutions are dependent on the environment and it cannot be deployed effectively over multiple scenarios. One of the major problems of nurse scheduling problem is the potential size of the set of individual schedules. Given that D is the number of days to be planned and any nurse can be assigned to S different shifts each day, then an order of DS schedules will be evaluated for a nurse. Most solutions using the integer/mixed generalized linear programming approach involves a linear relaxation of the master problem and an integer or mixed solution search procedure. In Jaumard et.al.(1998), the generalized model involves the generalization of the various conditions and constraints in nurse scheduling problem. Many critical conditions and constraints were listed in the paper such as the preference of allocation, allocation of specialty and others. In the subsequent section, we will expand on the constraints and conditions in Jaumard et. Al. to incorporate flexibility in nurse allocation by ward and skill specialty.

**The general model**

The general form of the nurse scheduling problem takes the form of either a maximization problem (in the case of nurse preferences) or a minimization problem (in the case of nurse costs). Before we formulate the objective function, let us first declare the various sets and parameters that will be used in the formulation.

Sets of Values

| | |
|---|---|
| S | Shifts of the day (1 – AM, 2- PM, 3 - MN) |
| N | Nurse |
| D | Days of the planning horizon |
| L | Leave |
| M | Demand |
| O | Supply |
| P | Preference |
| C | Cost |
| E | No of duties required |
| $S_1$ | Specific set $S_1$ |
| . | . |
| . | . |
| . | . |
| $S_n$ | Specific set $S_n$ |

Parameters

| | |
|---|---|
| $X_{nsd...}$ | Assignment for nurse n, shift s of day d (with $S_1, ... ,S_n$) |
| $P_{nd...}$ | Leave for nurse n for day d (with $S_1, ... ,S_n$) |
| $P_{nsd...}$ | Preference for nurse n, shift s of day d (with $S_1, ... ,S_n$) |
| $C_{nsd...}$ | Cost for nurse n, shift s of day d (with $S_1, ... ,S_n$) |
| $M_{sd...}$ | Demand for shift s of day d (with $S_1, ... ,S_n$) |
| $O_{sd...}$ | Supply for shift s of day d (with $S_1, ... ,S_n$) |
| $E_{ns...}$ | No of shifts for nurse n for shift s (with $S_1, ... ,S_n$) |

Thus the objective function can be declared as below

$$Maximize \quad Total\ Utility = \sum_{1}^{n} \overset{...}{...} \sum_{1}^{s} \overset{...}{...} \sum_{1}^{d} \overset{...}{...} X_{n...sd} P_{n...sd}$$

or

$$\text{Minimize} \quad \text{Total Cost} = \sum_{1}^{n} \cdots \sum_{1}^{s} \cdots \sum_{1}^{d} \cdots X_{n\ldots sd} C_{n\ldots sd}$$

In the case of penalized objective functions, they can be declared in the following forms.

$$\text{Maximize} \quad \text{Total Utility} = \sum_{1}^{n} \cdots \sum_{1}^{s} \cdots \sum_{1}^{d} \cdots \{X_{n\ldots sd} P_{n\ldots sd} - X_{n\ldots sd} C_{n\ldots sd}\}$$

$$\text{Minimize} \quad \text{Total Cost} = \sum_{1}^{n} \cdots \sum_{1}^{s} \cdots \sum_{1}^{d} \cdots \{X_{n\ldots sd} C_{n\ldots sd} - X_{n\ldots sd} P_{n\ldots sd}\}$$

In all the cases above, the decision variable $X_{n\ldots sd}$ is defined as below.

$$X_{n\ldots sd} = \begin{cases} 1 & \text{Nurse } n \text{ works on shift } s \text{ on day } d \text{ (with S1, \ldots, Sn)} \\ 0 & \text{Otherwise} \end{cases}$$

The above formulation will be used in the subsequent sections which define the constraints of the model.

**Nursing policies**

Nurses of various specialty and seniority are needed for different types of ward. Any nurse schedule produced must satisfy the standard constraints such as regulations, requirements of the special wards and ensuring that the leave plans of the nurses are met. Other constraints may include avoidance of shifts which are assigned too close together (e.g. afternoon shift followed by morning shift the next day), nurses-ward gender mis-match. In selected literatures, there are discussions about the ability of the senior nurses covering the junior nurses (Dowsland, 1998; You, Yu and Lien, 2010), this discussion will be incorporated into the general model. Most of the nursing policies constraints are legal requirements or industry adopted standards.

The first key constraint is the shift constraint. Any nurse is allowed to work no more than a single shift per working day.

$$\sum_{1}^{s} X_{n\ldots sd} \leq 1 \quad \forall\, n, \ldots, d \quad (C1)$$

This constraint ensures that no nurse is forced to work multiple shifts in a day. If there are any nurse working multiple shifts in a day, than the nurse will not receive sufficient rest which is undesirable and potentially dangerous. At the same time, there is minimum number of rest days that any nurse is entitled to. This constraint is known as the minimum rest days constraint.

The number of rest days for a schedule of d days can be expressed as the complement of the maximum number of y working days for any period of d days.

$$\sum_{1}^{s}\sum_{1}^{d} X_{n...sd} \leq y \quad \forall n, ... \quad (C2)$$

The constraint enforces proper rest days for the nursing staffs which is necessary for their individual well being. However, this constraint alone is insufficient to enforce proper amount of rest as it only prescribes the number of days to rest. The schedule might arrange the work in such a way that the nurse have to work almost continuously within the d days period. To prevent this, an additional constraint is needed to limit the maximum consecutive number of working days. This constraint is known as the maximum consecutive work days constraint and is used together with the C2 constraint for enforcement of rest days.

Given any K consecutive work days within the d days period, there must be at least 1 rest day.

$$\sum_{1}^{s}\sum_{k}^{k+4} X_{n...sd} < k \quad \forall n, ... \text{ where } k \leq d - k \quad (C3)$$

This constraint is developed further should the consecutive work days comprises of night shifts. Under most legislations, after i consecutive night shifts, there must be at least 1 sleep day and h rest day (equivalent to 1+h rest days, but only the h rest days will be counted towards legally required rest day).

$$\sum_{j}^{j+i-1} X_{n...,3,d} + \sum_{j+i}^{j+h+i}\sum_{1}^{s} X_{n...sd} \leq j \quad \forall n, ... \text{ where } j \leq d - i - h \quad (C4)$$

Where $s = 3 \text{ represents night shift}$

This constraint manifests in two separate forms. The first form enforces the start of the mandatory rest day for continuous night shifts. The second form enforces the maximum number of continuous night shifts which is usually a variation of the C2 but applied to night shifts with additional rest day requirements.

Annual leaves and training leaves are usually mandatory and forms part of the hospital's policies on human resource management as well as personal development and well being. At the same time, the leave policies mandate that any leaves must be approved unless there are extenuating circumstances. The leave constraint is defined as below.

For nurse n who has applied for leave on day d,

$$\sum_{31}^{s} X_{n\ldots sd} \leq L_{n\ldots d} \qquad \forall\, n, d, \ldots s_1, \ldots, s_n \qquad (C5)$$

$$L_{n\ldots d} = \begin{cases} 0 & \text{Nurse } n \text{ takes leave on day } d \text{ (with S1, \ldots, Sn)} \\ 1 & \text{Otherwise} \end{cases}$$

Where $S_1, \ldots, S_n$ represent other peripheral requirements.

One special case of consecutive day constraint is the MN – AM constraint, which is also known as the night – morning shift problem. As the night shift belongs to the previous day and the morning shift in the new day, the previous constraints does not prevent this scenario from happening. As such consecutive shifts are not permissible legally; we need to have an additional constraint for this.

Consecutive shifts Night-Morning are not permitted in the schedule.

$$X_{n\ldots 1, d+1} + X_{n\ldots 3, d} \leq 1 \qquad \forall\, n, d, \ldots s_1, \ldots, s_n \qquad (C6)$$

Where $s = 3$ represents night shift and $s = 1$ represents morning shift

In most countries, the wards are required to be staffed with a certain number of nurses that is determined by the number of patient in the ward. Should the number of patients be very low, there are requirements that the wards are staffed with minimal staffs. The nurses will normally need to be trained in the specialty of the ward if it is a special class ward and there are allocated minimum number of nurse of certain ranks. The allocation of nurses can be affected by other factors such as gender as well as language proficiency especially for hospitals with international patients. To simplify the requirements, we will generalize the requirement constraints.

The number of nurses in a ward must satisfy the following constraint,

$$\sum_{1}^{n} X_{n\ldots sd} = O_{s\ldots d} \qquad \forall\, n, s, \ldots s_1, \ldots, s_n \qquad (C7)$$

$$O_{s\ldots d} \geq M_{s\ldots d} \qquad \forall\, s, d, \ldots s_1, \ldots, s_n \qquad (C8)$$

Where $S_1, \ldots, S_n$ represent other peripheral requirements.

Because senior nurses with specialty training can operate as normal nurses, in certain cases, C8 may be modified slightly to carter for nurses with specialty training and rank to be deployed to other wards to cover shortages of nurses.

$$\{O_{s...t+1,w+1,d} - M_{s...t+1,w+1,d}\} + O_{s...twd} \geq M_{s...twd} \qquad \forall s,d,t,w,...s_1,...,s_n \qquad (C9)$$

Where nurses with t+1 specialty in ward w+1 can be transferred to work in ward w with specialty t nurses. The last key constraint is the number of shifts that a nurse needs to work. There are specified numbers of shifts that a nurse has to work and they need to be satisfied.

$$\sum_{1}^{d} X_{n...sd} = E_{ns...} \qquad \forall n,...,s \qquad (C10)$$

All the constraints discussed so far are mandatory in most hospitals as they are prescribed by the government as well as the industry standards. In the next section, we will discuss about the other types of constraints which are not mandatory but helps to improve the usability of the schedule.

**Hospital and Nurse Preferences**

Certain constraints are good to have but not mandatory. Most of the constraints are developed to meet the preferences of the nurses as well as maintaining the operational level of the wards. The hospitals do not wish to have schedules with excessive number of rest days which impacts the number of nurses that can be activated without concerns on rest days. On the other hand, most nurses do not enjoy the situation where they have to report to work in the morning for the next day after working afternoon shifts for the current day. The constraints below are the most common ones.

1. An afternoon shift should not be followed by a morning shift the next day.
2. Three or four consecutive night shifts are not preferred.

Managing such constraints can be difficult as they are not always mandatory and often they are preferences. To manage these constraints, we will derive two versions of the constraints. The first version assumes that these are hard constraints while the second version considers them as penalties on the objective functions.

For the case of afternoon shift being followed by a morning shift, the hard constraints can be formulated as below.

$$X_{n...2,d} + X_{n...1,d+1} \leq 1 \qquad \forall n,d,...s_1,...,s_n \qquad (C11)$$

Where $s = 2\ represents\ afternoon\ shift$ and $s = 1\ represents\ morning\ shift$

The objective function can be formulated to the following to incorporate the penalty in the form of a soft constraint. Before, we can do that, we need to declare the penalty variable Z.

$Z_{kv...}$    Penalty for assignment result of k and v

Where $k = X_{n...2,d}$ and $v = X_{n...1,d+1}$

Using declaration, we can then impose the penalty on the objective function

$$Maximize \quad Total\ Utility = \sum_{1}^{n}...\sum_{1}^{s}...\sum_{1}^{d}...\{X_{n...sd}P_{n...sd} - X_{n...sd}Z_{kv}\}$$

$$Minimize \quad Total\ Cost = \sum_{1}^{n}...\sum_{1}^{s}...\sum_{1}^{d}...\{X_{n...sd}C_{n...sd} + X_{n...sd}Z_{kv}\}$$

For the case of reducing multiple night shifts, the hard constraints can be formulated as below.

$$\sum_{1}^{j} X_{n...,3,d} \leq j - u \qquad \forall\ n, d, ... s_1, ..., s_n \qquad (C12)$$

Where $j - u$ is the maximum number of night shifts which is considered acceptable and meets the preference of the nurses. The constraint can also be incorporated into the objective function as a penalty value. To achieve that, we need to declare the penalty variable Z.

$Z_{v1,v2...,vn}$    Penalty for assignment result of $v_1 ... v_n$

Where $v_1 = X_{n...3,1}$ and $v_n = X_{n...3,j}$

$$Maximize \quad Total\ Utility = \sum_{1}^{n}...\sum_{1}^{s}...\sum_{1}^{d}...\{X_{n...sd}P_{n...sd} - X_{n...sd}Z_{v1,...,vn}\}$$

$$Minimize \quad Total\ Cost = \sum_{1}^{n}...\sum_{1}^{s}...\sum_{1}^{d}...\{X_{n...sd}C_{n...sd} + X_{n...sd}Z_{v1,...,vn}\}$$

Where the penalty parameter takes the following form,

$$Z_{v1,v2...,vn} = \begin{cases} penalty\ value & where\ v_1 = X_{n...3,1} = 1 ....v_n = X_{n...3,j} = 1 \\ 0 & Otherwise \end{cases}$$

Given the generalization of the nursing model, we will prepare an example of a general ward in a hospital. We will demonstrate the applicability of the general model to the problem and how it can be implemented in AIMMS.

**Model building in AIMMS: A general ward example**

The ward requires a schedule involving 20 nurses over a 2-week period for the general care. In this example, we use the case of 20 nurses with rank requirements for a single ward for a 14-day schedule. The nursing requirement stipulates that the senior nurse can cover the duties of the junior nurse but no vice versa. All the nursing policies constraints defined in the earlier section are implemented. We will also incorporate the nurses' general preference in the calculation of the total utility.

Sets of Values

| | |
|---|---|
| S | Shifts of the day (1 – AM, 2- PM, 3 - MN) |
| N | Nurse |
| D | Days of the planning horizon |
| R | Rank of Nurse |
| L | Leave |
| M | Demand |
| O | Supply |
| P | Preference |
| C | Cost |

Parameters

| | |
|---|---|
| $X_{nrsd}$ | Assignment for nurse n, rank r and shift s of day d |
| $L_{nrd}$ | Leave for nurse n, rank r for day d |
| $P_{nrsd}$ | Preference for nurse n, rank r and shift s of day d |
| $C_{nrsd}$ | Cost for nurse n, rank r and shift s of day d |
| $M_{rsd}$ | Demand for rank r and shift s of day d |
| $O_{rsd}$ | Supply for rank r and shift s of day d |

Thus the objective function can be declared as below

$$Maximize \quad Total\ Utility = \sum_{1}^{n}\sum_{1}^{r}\sum_{1}^{s}\sum_{1}^{d} X_{nrsd} P_{nrsd}$$

In all the cases above, the decision variable $X_{nrsd}$ is defined as below.

$$X_{nrsd} = \begin{cases} 1 & \text{Nurse n of rank r works on shift s on day d} \\ 0 & \text{Otherwise} \end{cases}$$

Below are the various constraints that need to be satisfied. The number of permissible shift in a single day is 1.

$$\sum_{1}^{S} X_{nsd} \leq 1 \qquad \forall\, n, d \qquad (D1)$$

The maximum number of working days for a schedule of 14 days is 11 working days.

$$\sum_{1}^{S}\sum_{1}^{d} X_{nsd} \leq 11 \qquad \forall\, n \qquad (D2)$$

The number of rest days for a schedule of 5 days has a maximum number of 4 working days.

$$\sum_{1}^{s}\sum_{d}^{d+4} X_{nsd} \leq 4 \qquad \forall\, n, s \text{ and } d \leq 10 \qquad (D3)$$

Under the legislations, given 3 consecutive night shifts, there must be at least 1 sleep day and 1 rest day.

$$\sum_{j}^{j+2} X_{n,3,d} + \sum_{j+3}^{j++3}\sum_{1}^{s} X_{nsd} \leq 3 \qquad \forall\, n, \ldots \text{ where } j \leq 10 \qquad (D4)$$

The nurse cannot work on any days that she has applied for leave or assigned training leave.

$$\sum_{1}^{s} X_{nsd} \leq L_{nd} \qquad \forall\, n, d \in [1,14] \qquad (D5)$$

Consecutive shifts in the form Night-Morning shifts are not permitted in the schedule under the legislation.

$$X_{n,s_1,d+1} + X_{n,s_2,d} \leq 1 \qquad \forall\, n, d \in [1,13] \qquad (D6)$$

Where $s_1 = 1\, represents\, morning\, shift$ and $s_2 = 3\, represents\, night\, shift$
The number of nurses in general ward must satisfy the following constraint,

$$\sum_{1}^{n} X_{nrsd} = O_{srd} \qquad \forall\, n, s, r, d \qquad (D7)$$

$$O_{srd} \geq M_{srd} \qquad \forall\, s, d, r \qquad (D8)$$

As the senior nurses can cover the duties of the junior nurses, the following constraints can be used for the junior rank demand constraint.

$$\{O_{s,r_1,d} - M_{s,r_1,d}\} + O_{srd} \geq M_{srd} \qquad \forall\, s, d, r \qquad (D9)$$

All the nurses are required to fulfill the required number of shifts in a d (14 days) days period.

$$\sum_{1}^{d} X_{nsd} = E_{ns} \qquad \forall\, n, s \qquad (D10)$$

Using all the parameters, constraints as well as variables, we input all the information into AIMMS as set, parameters, constraints and objective functions. The optimization was done using an intel centrino duo laptop.

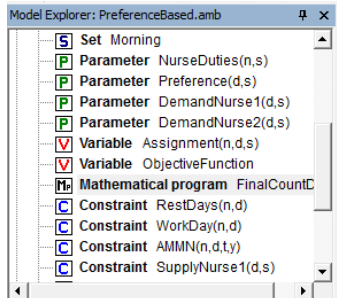

Screenshot 1: Setup of the variables, parameters and constraints.

Due to the nature of the formulation, the nursing scheduling problem is constructed in a linear integer programming form. In this case, it can be solved using CPLEX quite easily.

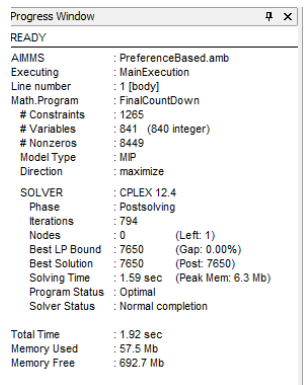

Screenshot 2: CPLEX Solver result.

From the results, the problem has been solved in 1.92 seconds which is extremely fast. At the same time, we can observe that solution is considered optimal by the software. Repeats of the optimization indicate that the solution can be found in less than 2 seconds for most of the time. AIMMS produces the result in a simple form that can be used by the nurses directly.

**Model building in AIMMS: Li et. Al., 2003**

The nurse scheduling problem as defined by Li et. Al. (2003) prepares a simple schedule involving 27 nurses over a 1-week period for the general care. In this example, there are neither rank requirements nor ward requirements for a 7-day schedule. The nursing requirement stipulates that there can only be 1 night shift with a maximum of shifts in the 7 days period. Due to the nature of the problem, only certain applicable nursing policies constraints defined in the earlier section are implemented. The problem is formulated as cost problem as opposed to a utility problem. Each nurse has a cost attached to them which can be reduced by assigning them to duties which they have indicated their preferences.

Sets of Values

| | |
|---|---|
| S | Shifts of the day (1 – AM, 2- PM, 3 - MN) |
| N | Nurse (27) |
| D | Days of the planning horizon (7) |
| L | Leave |
| M | Demand |
| O | Supply |
| P | Assigned Cost |
| C | Cost |

Parameters

| | |
|---|---|
| $X_{nsd}$ | Assignment for nurse n, shift s of day d |
| $L_{nd}$ | Leave for nurse n for day d |
| $P_n$ | Assigned Cost for nurse n |
| $C_{nsd}$ | Cost for nurse n, shift s of day d |
| $M_{sd}$ | Demand for shift s of day d |
| $O_{sd}$ | Supply for shift s of day d |

Thus the objective function which involves cost is declared as below

$$Minimize \quad Total\ Cost = \sum_{1}^{n} P_n - \sum_{1}^{n}\sum_{1}^{d}\sum_{1}^{s} X_{nsd} C_{nsd}$$

In all the cases above, the decision variable $X_{nrsd}$ is defined as below.

$$X_{nsd} = \begin{cases} 1 & \text{Nurse } n \text{ works on shift } s \text{ on day } d \\ 0 & \text{Otherwise} \end{cases}$$

The number of permissible shift in a single day is 1.

$$\sum_{1}^{S} X_{nsd} \leq 1 \quad \forall\, n, d \quad (D1)$$

The maximum number of working days for a schedule of 7 days is 5 working days.

$$\sum_{1}^{S}\sum_{1}^{d} X_{nsd} \leq 5 \quad \forall\, n \quad (D2a)$$

The maximum number of working night shifts for a schedule of 7 days is 1 days.

$$\sum_{1}^{d} X_{n,3,d} \leq 1 \quad \forall\, n \quad (D2b)$$

Where $s = 3\ represents\ nigt\ shift$

The nurse cannot work on any days that she has applied for leave or assigned training leave.

$$\sum_{1}^{S} X_{nsd} \leq L_{nd} \quad \forall\, n, d \in [1,7] \quad (D3)$$

Consecutive shifts in the form Night-Morning shifts are not permitted in the schedule under the legislation.

$$X_{n,s_1,d+1} + X_{n,s_2,d} \leq 1 \quad \forall\, n, d \in [1,6] \quad (D4)$$

Where $s_1 = 1\ represents\ morning\ shift$ and $s_2 = 3\ represents\ night\ shift$

The number of nurses in general ward must satisfy the following constraint,

$$\sum_{1}^{n} X_{nsd} = O_{sd} \quad \forall\, n, s, d \quad (D5)$$

$$O_{sd} \geq M_{sd} \quad \forall\, s, d \quad (D6)$$

Where $M_{1,d} = 6, M_{2,d} = 6, M_{3,d} = 3$

Using the information, we input all the information into AIMMS as set, parameters, constraints and objective functions. The optimization was done using an intel centrino duo.

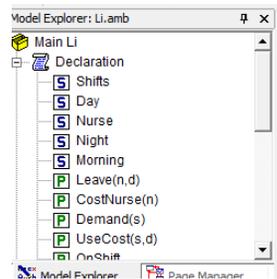
Screenshot 1: Setup of the variables, parameters and constraints.

The nursing scheduling problem is constructed in a linear integer programming form which can be solved using MOSEK quite easily.

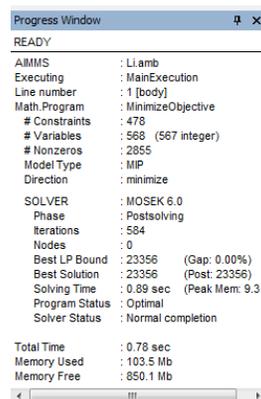
Screenshot 2: MOSEK Solver result.

From the results, the problem has been solved in 0.89 seconds which is extremely fast. However the solution is considered sub-optimal as compared to the results produced by Roster Booster which contains the original implementation of the solution (Li et. Al., 2003). This may be due to the nature of the problem which is better served by the algorithm designed to solve the problem. Repeats of the optimization indicate that the solution can be found in less than 2 seconds for most of the time.

**Conclusion and future directions**

We have developed a flexible linear integer programming model for nurse scheduling. The model incorporated the general constraints required. The resulting schedule includes balanced schedules in terms of the distribution of shift duties, fairness in terms of the number of consecutive night duties and the preferences of the nurses. This is an improvement over the traditional manual approach which is costly in terms of labor as well as inefficient in producing a good schedule.

The flexible linear integer programming model allows for simple implementation in mathematical programming program easily. Due to the formulation of the problem, the linear nature allows for the application of solvers which are extremely efficient at solving such problems. Given that the nurse scheduling problem is considered by many to be a NP hard problem, the ability to solve it using a linear mathematical solver makes it easier to implement an automated approach. In this regard, the model makes it a practical computerized tool. The model is also generalized to a degree where it is possible to extend it to incorporate various new constraints and preferences.

The future direction of this work involves the building and solving of the model using open source solvers which makes it easier for the hospitals to use. At the same time, there are certain requirements and constraints which might not have been captured in the constraints developed. With the evolving legal frameworks, there remain the possibilities of new legal requirements for the nurse scheduling problem which will need to be formulated to incorporate into the model. The current model also considered only two major preferences of the hospital and nurses which might not be the case in other circumstances given the variations in human preferences. Nevertheless, the general form has made provisions for the extension of the model should the need ever arises.